\documentclass{acm_proc_article-sp}
\usepackage{times}
\usepackage{helvet}
\usepackage{courier}
\usepackage{graphicx}
\usepackage{algorithmic}
\usepackage{algorithm}
\usepackage{ amsmath}

 \newcommand{\remove}[1]{}    
\begin{document}
%
\title{Information Relaxation is Ultradiffusive}

\numberofauthors{2}
%


\author{
\alignauthor
Rumi Ghosh\\
\affaddr{Social Computing Lab}\\
\affaddr{Hewlett Packard Research Labs}\\
\affaddr{Palo Alto, California, USA}\\
\email{rumi.ghosh@hp.com}
\alignauthor
Bernardo A. Huberman\\
\affaddr{Social Computing Lab}\\
\affaddr{Hewlett Packard Research Labs}\\
\affaddr{Palo Alto, California, USA}\\
\email{bernardo.huberman@hp.com}
}
\maketitle
\begin{abstract}

We investigate how the overall response to a piece of information (a story or an article) evolves and relaxes as a function of time in social networks like Reddit, Digg and Youtube. This response or popularity is measured in terms of the number of votes/comments that the story (or article) accrued over time. We find that the temporal evolution of popularity can be described by a universal function whose parameters depend upon the system under consideration.  Unlike most previous studies, which empirically investigated the dynamics of voting behavior, we also give a theoretical interpretation of the observed behavior using ultradiffusion. 

Whether it is the inter-arrival time between two consecutive votes on a story on Reddit or the comments on a video shared on Youtube, there is always a hierarchy of time scales in information propagation. One vote/comment might occur almost simultaneously with the previous, whereas another vote/comment might occur hours after the preceding  one.  This hierarchy of time scales leads us to believe that  the dynamical response of users to information is ultradiffusive in nature.
We show that a ultradiffusion based stochastic process can be used to rationalize the observed temporal evolution.



We demonstrate that the results predicted by simulating the ultradiffusion process are in close correspondence to the actual observations. 
\end{abstract}
\section{Introduction}

Recent years have seen an exponential increase in the popularity and growth of online social networks.
Although these networks have not yet replaced the traditional sources of information like the news websites, they are fast becoming alternate conduits of information. Be it the US airplane crash in the Hudson River in 2009\footnote{http://jalcommunication.com/hudson-river-plane-crash-demonstrates-power-of-twitter/}, the Arab Spring \footnote{http://smallwarsjournal.com/jrnl/art/what-the-arab-spring-tells-us-about-the-future-of-social-media-in-revolutionary-movements},  the floods in Thailand in 2011 \footnote{http://www.techinasia.com/thailand-flood-social-media-innovation/}, the Hurricane Sandy in 2012 \footnote{http://www.theguardian.com/world/us-news-blog/2013/feb/20/mta-conedison-hurricane-sandy-social-media-week}
 or the Boston marathon bombing \footnote{http://www.voanews.com/content/multi-social-media-play-huge-role-in-solving-boston-bombing/1649774.html} or the recent bus -train crash in Ottawa in 2013 \footnote{http://www.huffingtonpost.ca/2013/09/18/oc-transpo-via-rail-crash-ottawa\_n\_3948062.html}, social media has started playing an increasingly important role in the dissemination of information and crisis management. The propensity of viral spread of information through these social networks, have led to their growing and successful use in marketing  campaigns like Nike's ``finding greatness" campaign \footnote{http://www.copywriterjournalist.com/2012/08/05/nike-find-your-greatness-campaign-videos-london-2012-olympics/} and also electoral   campaigns like President Barrack Obama's presidential campaigns in United States. Their effectiveness in spreading information have made them powerful mouthpieces for stars and starlets and have created their own league of celebrities like Justin Bieber and Lady Gaga. Thus, understanding the dynamics of information flow in these networks is crucial to characterize, predict, improve and innovate upon their role as useful information channels. 

There has been a lot of work to understand information flow on networks. 
The dynamics of information flow can studied in relation to the users participating in it and the network that they create. Work in this direction includes the study of information spread such as characterization of information propagation in the blogosphere\cite{Gruhl04},  the characteristics of information content that gets viral(trends) \cite{Kwak10},  nature of propagation of  recommendations \cite{viraldynamics}, role of social networks in information spread \cite{BakshysocN} and so on.  The role of individuals in a network  in the spread of information and their ability to influence others has also been actively studied.  This includes investigations on the probability of a person to influence people within ones social circle as compared to those outside it \cite{wu04information}, the comparative analysis of different methods to measuring influence during information propagation \cite{Cha2010} to name a few.  Often, the underlying network through which information spreads is not known and there have been attempts to infer near-optimal networks that might lead to the observed contagion. Unlike \cite{GomezRodriguez13}, the objective of this paper is \emph{not} to infer the structure of the social network responsible for the observed observed temporal dynamics.  

The focus is the study of the temporal evolution of user response to information and factors that can rationalize the observed  behavior. The response to information is often in terms of votes/comments. The number of comments/votes accrued over time gives a measure of popularity of that piece of information with time. Like \cite{lermanGhosh}, we empirically investigate this temporal evolution of popularity of stories in different social networks.   However, unlike most of the previous work in this direction, we also provide a theoretical interpretation of the observed phenomena. The observations could be explained using  hierarchy of time scales and ultradiffusion. 
Using ultradiffusion to understand the temporal phenomena of popularity growth gives us  invaluable insights into the relaxation of information response. It can serve as a powerful analytic tool can complement work that attempts to explain information spread based on the underlying social network.





There exists a hierarchy of time scales for information spread through an online social network. On one hand, some user might rebroadcast information immediately when he receives it. On the other hand, some other user might wait for a long time before rebroadcasting it. Yet another user, might wait for majority of his friends to rebroadcast an information before rebroadcasting it himself. One vote or rebroadcast can occur almost immediately after the previous. Another rebroadcast might occur a few minutes after the previous one. Further, another broadcast might occur hours after the one preceding it. 
Some other examples of hierarchical systems include taxonomy of objects, places, concepts, events, properties, and relationships; structure of organisms having atomic cells as basic building blocks; ontology in information science so on and so forth.  A common feature of hierarchical systems   is the existence of many timescales leading to anomalous dynamics of relaxation. Physicists have described the nature of dynamics and modeled diffusive processes through these systems  \cite{BachasHuberman, Kerzberg} with the aid of \emph{ultrametricity} and \emph{ultradiffusion}. 
Therefore, due to the hierarchical time scales of information response through online social media, we postulate that ultradiffusion can also be used describe the nature of dynamics of information  response and relaxation over time.



We study the temporal evolution of information response in social media like Twitter, Reddit and Youtube. 
We are able to quantify the temporal evolution of rebroadcasts of information using different measures of rebroadcast such as votes or comments change with time.
We observe that the probability of rebroadcasts or the popularity growth follow a universal function irrespective of the social media studied. The parameters of the function can be learned from the data under consideration. 
In contrast to the Poisson process, ultradiffusion can successfully explain the exponential nature of the growth curve observed. 

Ultradiffusion shows that the state space of the associated stochastic process has a hierarchical structure and the  observations are a consequence of transitions within this hierarchical structure. Based on the temporal patterns of popularity, we can make powerful deductions about the the underlying hierarchical structure and the ultradiffusion process. For example, if the observed temporal trend is exponential, we can deduce that the underlying hierarchical structure is finite and learn its size and the parameters of the ultradiffusion process. 




Ultradiffusion is closely associated with the notion of ultrametric distance. 
 Along with  common distance metrics include the Euclidean and Manhattan distance, it is another important though rarely used metric for distance or "closeness". 
 One of our key contributions in this paper is to show the ultrametricity of information response.
We show that the distance between any two rebroadcasts (votes or comments) is ultrametric.
Ultradiffusion can then simulated using a transitions in the state space of the underlying stochastic process where the transitions are a function of  this distance.  

 We demonstrate that that the probability of rebroadcasting (or voting) predicted by this process very closely correlated with the observed probability of rebroadcasting irrespective of the social media considered. Our model seems to describe dynamics of voting on stories in Digg and Reddit, the dynamics of commenting on videos on Youtube with high accuracy. It also seems to satisfactorily explain the evolution of downloads from a web server.

We also show that using ultradiffusion to explain phenomena containing hierarchy of time scales is not just limited to social networks. The spread of popularity of a research paper publicized in the news media and measured by the number of downloads with time can also be explained using ultradiffusion.

In the next Section we discuss related work. 
Following that we  introduce ultrametricty and ultradiffusion and the datasets under consideration. Next we empirically characterize response to information and its relaxation. 
We show how the distance between any two rebroadcasts of a piece of information is ultrametric. We create a transition matrix whose rates are defined by the ultrametric distance and is used to   model ultradiffusion. Ultradiffusion is  then used to explain the behavior observed. In the section on experiments, we present the experiments carried out, followed by the results and evaluation. We conclude with discussion and future work.

\section{Related Work}
\label{sec:related}
Reference \cite{Watts} define a class of searchable networks. Their definition of searchablity is also based on the assumption of hierarchy of the social space be it an organization or community. Each individual is a leaf of this hierarchical tree. The distance between two people is assumed to be ultrametric and is taken distance from their common ancestor.  We on the other hand claim that observed dynamics  like voting on a story or sharing of a video could be best explained using ultrametricity and ultradiffusion. An underlying ultrametric structure of the social network could be one of the many potential causes for the ultrametricity in dynamics.

A lot of empirical work has been done on the nature of spread of information. For example \cite{wu04information} showed that a story relevant to an individual is more likely to be relevant to others within his social circle than to those outside it. \cite{viraldynamics} study how recommendations spread in a person-person network and model it using a simple stochastic process. 
\cite{lermanGhosh} do an empirical analysis of information flow in online social networks of Digg and Twitter. On the other hand \cite{Cha2010, Romero2011} measure influence during information flow on Twitter. Similarly, \cite{Kwak10} did another quantitative study of information diffusion in Twitter and showed that majority of stories diffusing (or trending topics) are headlines or persistent news. \cite{GomezRodriguez13} try to infer the structure of the network based on diffusion data. 

Spreading processes have also been extensively studied in epidemiology. 
Models like SIS (susceptible-infectious-susceptible) and SIR (susceptible-infectious-recovered) have often been used to model the spread of diseases \cite{Bailey}. However these models assume that the population is homogeneous. To get closer to real life, \cite{Hethcote} divided the population into categories based on age, sex and so on, and then treated the subcategories as homogeneous. \cite{Kephart} studied the spread of computer viruses. To take the heterogeneity into account, they represented interactions between nodes as a directed graph leading to a single mean-field equation.
\cite{Vespignani2001} relaxed the homogeneity assumption further by assuming the structure of the network to be scale-free and proposing the heterogeneous mean-field (HMF) theory to study the spread of diseases through them. 
 They claimed that the epidemic threshold for a scale-free graph is zero implying that even for arbritarily low transmission probabilities, a finite portion of the graph is infected. \cite{Prakash} conjecture that for any virus propagation model, the epidemic threshold depends on the largest eigenvalue of the adjacency matrix of the network. However, \cite{Castellano} claim that while this may hold true for SIS models, the HMF predictions are much for accurate for SIR networks. \cite{NewmanEpidemic} showed that a power-law degree distribution with an exponential cutoff would lead to a non-zero epidemic threshold. However, this threshold would still be very close to zero.

\cite{wu04information} argued that  information flow might not be analogous to viral epidemics. They stated that while the claim of an  zero (or close to zero) epidemic threshold in HMF predictions\cite{Vespignani2001, NewmanEpidemic} might be valid for viral spread of diseases (which spread indiscriminately) information flow might be different. This is because in information spread, the host might be selective and pass information only to people interested in it.  \cite{weng:virality,Backstrom06,CentolaD} show that information spread is affected by the structure of the network and clustering within it.  \cite{Versteeg11icwsm} also demonstrate that in epidemic models  like HMF, predictions  do not correctly describe the spreading dynamics in real social networks like Digg. In stead they propose a different contagion model, the friend saturation model to describe the observed behavior. \cite{Cui13} propose Orthogonal Sparse LOgistic Regression  method to predict epidemic outbreaks.

All of these studies, investigate or try to understand some empirical property, or structure of information flow and the associated properties as a function of the underlying social network.  On the other hand, like \cite{Gabor}  we attempt to understand the response to information as a function of time. \cite{Gabor} show that initial evolution of votes on stories in Digg and Youtube could be used to forecast their future popularity. We on the other hand study the temporal  evolution pattern of rebroadcasts (votes or comments) on Reddit, Digg and Youtube with time and show that it can be modeled using ultradiffusion.


While there has been work \cite{Malmgren08,Barabasi05nature} on the inter-arrival times between emails (events), we focus on the evolution of a event occurrences or response to information with time.
\section{ Ultrametricity and Ultradiffusion}
One of the important characterization of information is the relaxation of its response or its popularity trace over time. Here, we provide an in depth empirical, experimental and theoretical analysis  of this trace.
If a social news aggregator is considered and the information is a news story, its popularity with time is measured by the number of votes it collects over time. Similarly, if the social media is a video sharing website and the information propagating is a video shared, the observed phenomenon indicating change in popularity might be the number of comments that the video receives with time. We take some real-life examples of temporal evolution of response to information. We consider the growth of popularity  of stories on Reddit, news on Digg and videos on Youtube. We also study the number of downloads of a research paper after it was widely publicized \cite{Sornette}.The  quantifiable measure of response to information in the four cases considered are: the increase in the number of votes on Digg and Reddit stories over time, the temporal increase in the number of comments on Youtube videos and the number of downloads from a university server with time.  We define the observable response (voting/ commenting/downloading) as the \emph{rebroadcast phenomena}. Each vote/comment/download is an \emph{event}. The temporal evolution of the rebroadcast phenomena is called  the \emph{counting process}. 

\subsection{Ultrametricity}
The observed counting process could  perhaps be rationalized as  ultrafiffusive if it has signatures associated with an ultradiffusive stochastic  process.
Ultradiffusion is a stochastic process and like any stochastic process comprises of a state space $S$ which is a collection of random variables $(X_t, t\in T)$ where $t$ represents time, $X_t$ is the random variable associated with a rebroadcast/ event at time $t$.  The collection of timestamps at which event occurrences are considered is $T$. $X_t$ is $1$ when an event occurs and is $0$ otherwise.  

Unlike Poisson process which assumes that the event occurrences are independent of each other, ultradiffusion elicits that a later event might be caused by or correlated to an earlier event or a combination of earlier events. The earlier event  in turn might be independent or it might be correlated to a combination of even earlier events. This leads to a hierarchical causal/correlational model of prior event occurrences which can be used to predict the occurrence of a new event. In other words, ultradiffusion proposes that the observed pattern of events is a consequence of an  underlying hierarchy of states. 

In this hierarchical model, an event temporally nearer to the occurring event has a greater probability of affecting it. In other words the correlation between two events is determined by a notion of "closeness" or distance between them. Ultradiffusion hypothesizes that this distance between two event occurrences is ultrametric.
A set of states $X_t \in S$ form an  \emph{ultrametric space} and function $d(X_{t_i},X_{t_j})\to R$ is a \emph{ ultrametric distance metric} if:
\begin{enumerate}
\item $d(X_{t_i},X_{t_j}) \ge 0$ (non-negative)
\item  $d(X_{t_i},X_{t_j})=0$ if $i=j$ 
\item $d(X_{t_i},X_{t_j})=d(X_{t_j},X_{t_i})$ (symmetry)
\item $d(X_{t_i},X_{t_j})\le max(d(X_{t_i},X_{t_k}),d(X_{t_k},X_{t_j}))$ (\emph{ultrametric property i.e.} the distance between any arbitrary events  $X_{t_i}$ and $X_{t_j}$ is less than or equal to the maximum of the  distance between state $X_{t_i}$ and $X_{t_k}$ (any other state) and that between $X_{t_k}$ and $X_{t_j}$).

\end{enumerate}




\subsection{Ultradiffusion}
Ultradiffusion is best studied by resorting to the master equation with any two states $X_{t_i}, X_{t_j}$ describing the evolution of probability over time \cite{BachasHuberman}:
\begin{equation}
\frac{dP_{X_{t_i}}}{dt}=\sum_{j=1}^{N} \epsilon_{X_{t_i} X_{t_j}}P_{X_{t_j}}
\label{eq:diffusion}
\end{equation}
 Here $\epsilon$ is transition matrix, whose element corresponding to states $X_{t_i}$ and $X_{t_j}$,  $\epsilon_{X_{t_i} X_{t_j}}$ satisfies the ultrametric property:
\begin{equation}
\epsilon_{X_{t_i} X_{t_j}} \ge min{(\epsilon_{X_{t_i} X_{t_k}} ,\epsilon_{X_{t_k} X_{t_j}} )}
\end{equation}
for distinct $i$,$j$ and $k$. Here $N$ is the number of states. The conservation of probability requirement entails that for $i=j$ i.e. the diagonal elements of the transition matrix, $\epsilon_{X_{t_i} X_{t_i}}=-\sum_{i \ne j}\epsilon_{X_{t_i} X_{t_j}}$.

\remove
{
Solving for $P$ gives us 
\begin{equation}
P(t)=e^{(\epsilon t)}P(0)
\label{eq:diffusion1}
\end{equation}
where P(0) is the starting vector.
}
The transition matrix $\epsilon$ can be any monotonically decreasing function of the ultrametric distance function $d(X_{t_i}, X_{t_j})$ for distinct  states $X_{t_i}$ and $X_{t_j}$.

Without loss of generality we take $\epsilon_{X_{t_i}, X_{t_j}}=e^{(-\mu d(X_{t_i}, X_{t_j}))}$ for $i\ne j$ where $\mu$ is the scaling function. 

The autocorrelation  $P_{X_{t_i}}(t)$ is the probability of finding the system at the initial state $X_{t_i}$ after time $t$.  Bachas and Huberman \cite{BachasHuberman} have shown that the exact solution for the autocorrelation function associated with Equation \ref{eq:diffusion} (for an ultrametric space defined by a hierarchical tree) is of the form :
\begin{equation}
P_{X_i}(t)=\alpha+\sum \beta_n e^{-\gamma_n t}
\label{eq:4}
\end{equation}
The constants $\alpha$, $\beta_n$ and $\gamma_n$ can be derived analytically  as shown in Equation \ref{eq:5} in the Appendix.
When the number of states is finite, the resultant autocorrelation function is exponential in nature. Otherwise, it follows a power law as shown in the Appendix. Next, we describe the dataset studied.

\section{Datasets and Measurements}
\label{sec:data}
We  study data from social networking sites, Digg, Youtube and Reddit.  We also investigate the evolution of downloads from a web server \cite{Sornette}.

\subsection{Reddit}
Reddit \footnote{www.reddit.com} is a social news and entertainment website. It has a homepage and other pages like "rising","controversial","new" where stories are ranked based on different criteria.  Registered users can post a links to hyper-linked stories/news elsewhere or some can write their own ("self") stories/news in text format. Other users can vote in favor of or opposing these submissions.
These votes determine the ranking of the story in the different ordered lists/pages of stories hosted by this website. The data \footnote{http://www.reddit.com/r/redditdev/comments/bubhl/ csv\_dump\_of\_reddit\_voting\_data/} we used has the votes (and the timestamps of these votes) on more than 16000 links in Reddit.
There were 31,927 unique users and more than 7 million votes. Of these 9700 stories/links having more than 50 votes which comprise our dataset.
As expected the distribution of votes over stories follows a power law distribution, with many stories receiving few votes and few stories receiving many votes. We study, the temporal evolution of votes in each of these stories and on average.
\begin{figure}[htbp] %
   \centering
   \begin{tabular}{c}
 \includegraphics[width=0.5\textwidth]{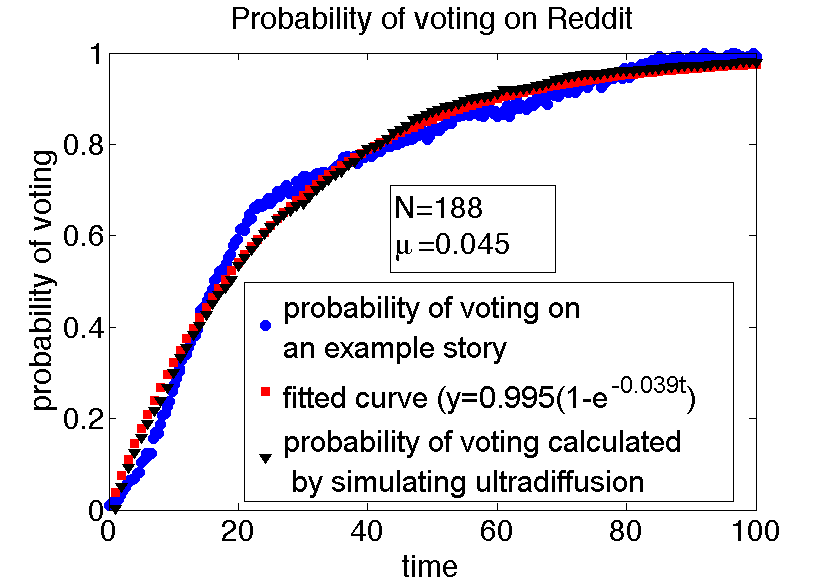}  \\
(a)\\
 \includegraphics[width=0.5\textwidth]{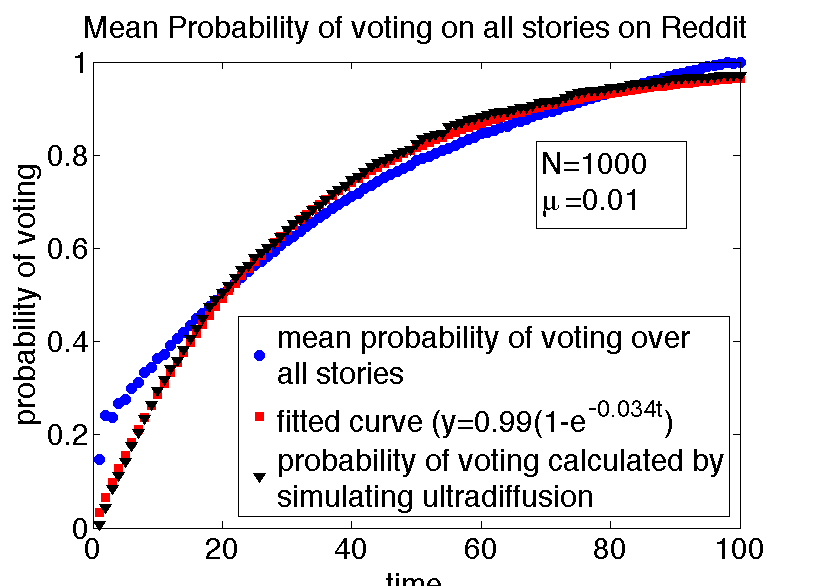}\\
(b)
   \end{tabular}
   \caption{(a)The observed probability of  voting on Reddit with time for a sample story (in blue). The exponentially fitted curve is in red ($R^2=0.98$) and the calculated probability of voting from a simulated ultradiffusion process is shown in black ($R^2=0.98$). (b)The blue curve gives the observed mean probability of voting on Reddit. The exponential curve fitting is given in red ($R^2=0.96$). The mean probability of voting generated by an ultradiffusion simulation is given in black ($R^2=0.93$).}
   \label{fig:Reddit}
\end{figure}

\subsection{Digg}
Digg\footnote{www.Digg.com} is a social news aggregator, which like Reddit, allows users to posts links to news stories. When a story is submitted by a user, other users can vote on it. The number of diggs or votes received by a story shows how popular a story is and plays an important role in its ranking with respect to other stories on the front page of Digg. The temporal voting behavior on a story is indicative of how the story propagates amongst the users of Digg with time.
The dataset we used \cite{lermanGhosh} comprises of voting on stories promoted to Digg's front page over a period of a month in 2009 (July). It contains  355K  votes on 2060 popular stories made by 100K distinct users. \\

\begin{figure}[htbp] %
   \centering
   \begin{tabular}{c}
 \includegraphics[width=0.5\textwidth]{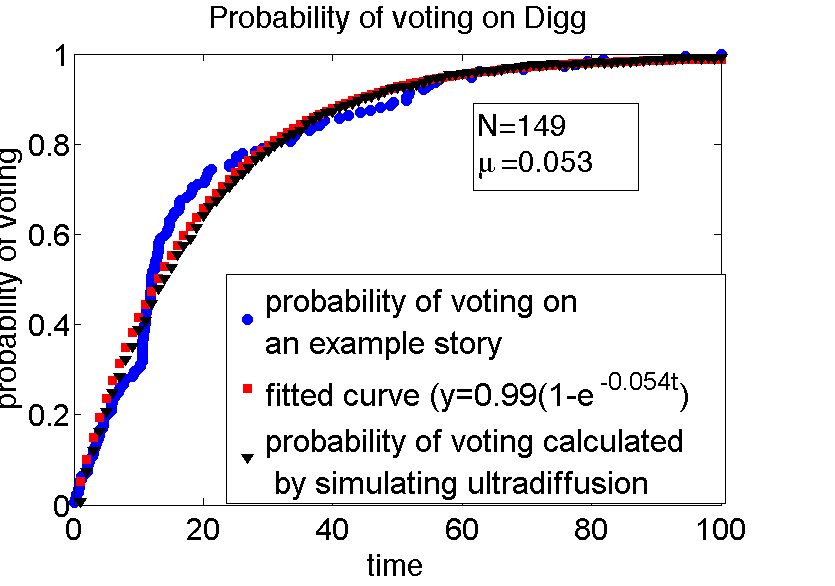} \\
(a)\\
 \includegraphics[width=0.5\textwidth]{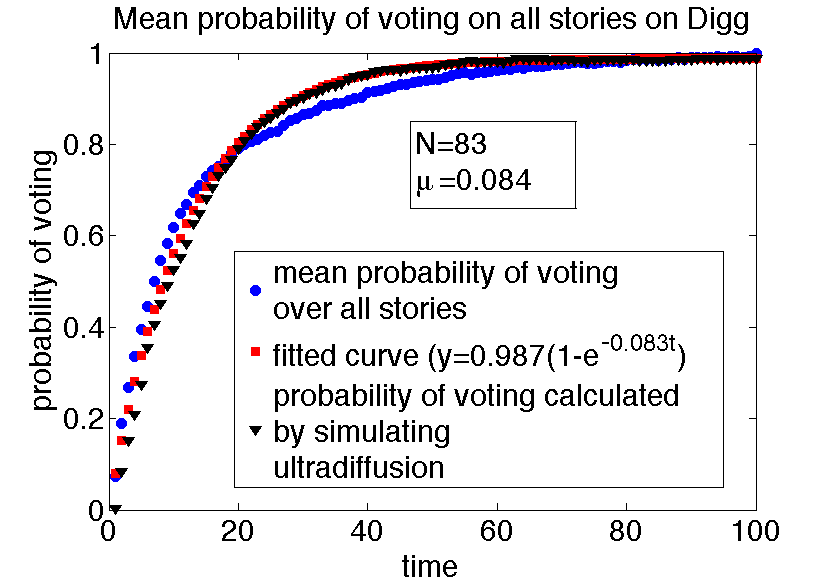}   \\
(b)
   \end{tabular}
   \caption{(a)The probability of voting on Digg with time for a sample story. The observed probability of voting is in blue. The fitted exponential curve is in red ($R^2=0.976$) and the probability of voting obtained by simulating ultradiffusion is in black ($R^2=0.976$) (b)The average probability of  voting over all stories is shown in blue and the fitted curve is shown in red ($R^2=0.97$). The average probability of voting simulated using ultradiffusion is in red ($R^2=0.95$). }
   \label{fig:Digg}
\end{figure}

\subsection{Youtube}
Youtube\footnote{www.youtube.com} is a popular video sharing platform. An individual can upload a video on Youtube others can view it. Often some videos receive word-of-mouth appreciation and spread through friendship networks and become viral. For example, one of the viral videos in 2012 "Gangnam Style" reached over a billion views \footnote{http://articles.latimes.com/2012/dec/17/business/la-fi-ct-viral-video-top-ten-20121218}. People also have the option of commenting on these videos. The commenting behavior is the quantifiable measure of the change in information popularity over time.
\begin{figure}[htbp] %
   \centering
   \begin{tabular}{cc}
 \includegraphics[width=0.5\textwidth]{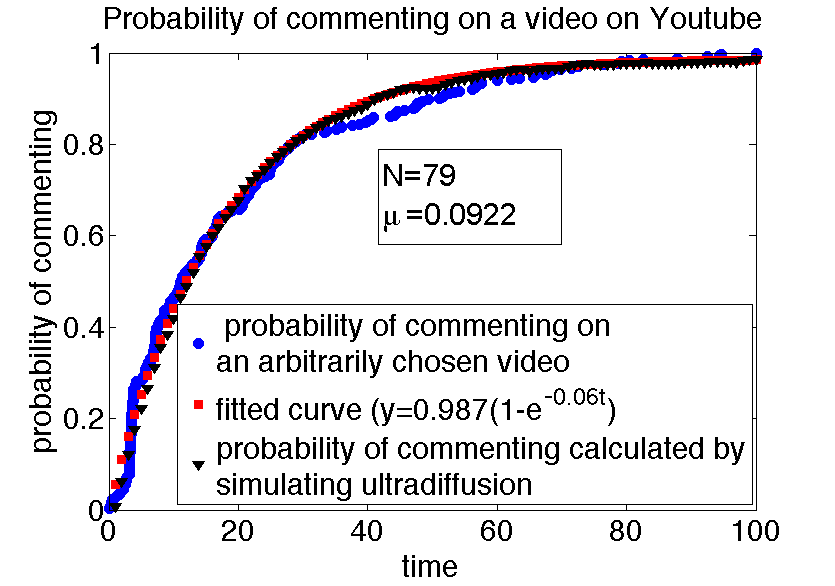} \\
(a)\\
 \includegraphics[width=0.5\textwidth]{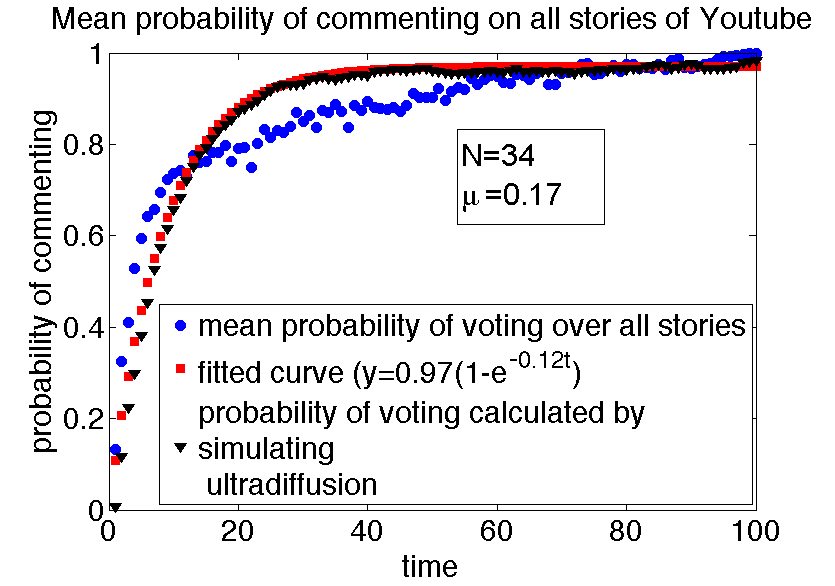}   \\
(b)
   \end{tabular}
   \caption{(a) The probability of  commenting on a sample story Youtube. The curve in blue represents the observed probability of commenting. The red curve shows the fitted curve ($R^2=0.986$) and the probability of commenting obtained by simulating ultradiffusion is shown in black ($R^2=0.99$). (b)  The mean probability of commenting is shown in blue. The fitted curve is in red ($R^2=0.80$). The probability of commenting derived from ultradiffusion simulation is in black ($R^2=0.80$)}
   \label{fig:Youtube}
\end{figure}
We extracted the comments and the timestamps at which these comments were posted for more than 20 videos  of advertisements in 2012 and 2013 on Youtube having more than 7500 comments. 
We study the evolution of the number of comments accrued by these videos and model it using ultradiffusion. 
\subsection{Server Downloads}
Johansen et al.\cite{Sornette} studied the response of people to new information. They investigated and modeled the number of downloads of a research paper from an authors webpage in Niels Bohr Institute. This dataset was collected after the paper was publicized by a Danish newspaper. We show here that ultradiffusion gives an alternate (possibly better) explanation of the observed variation of downloading activity with time.
\begin{figure}[htbp] %
   \centering
   \begin{tabular}{c}
 \includegraphics[width=0.5\textwidth]{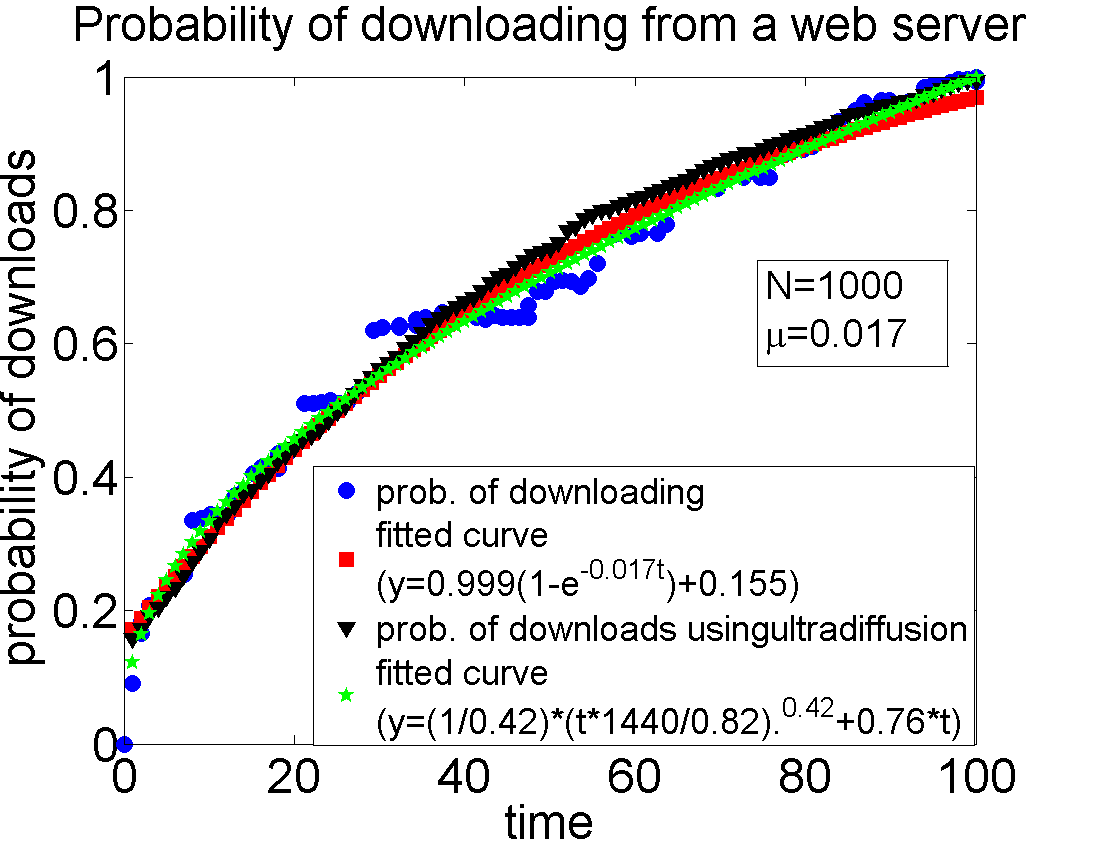}   \\
   \end{tabular}
   \caption{The probability of downloads of a paper from Niels Bohr Insititute server with time is shown in blue \cite{Sornette}. The curve in green ($y=(1/0.42)*(t*1440/0.82).^{0.42}+0.76*t$)is one postulated to describe the phenomena in \cite{Sornette} ($R^2=0.988$). The curve shown in red ($R^2=0.98$) is the fitting of the universal function $y=h_1(1-e^{-h_2 t})+h_3$ ($h_1=0.999, h_2=0.017, h_3=0.155$). The black triangles show the evolution of downloads simulated by ultradiffusion($R^2=0.97$). }
   \label{fig:server}
\end{figure}

\subsection{Measurements and Observations }
Like \cite{lermanGhosh}, we observe that most stories/videos in most social media accrue a lot of new votes/comments initially after being posted. Gradually the number of new votes/comments slow down with time and finally it reaches a saturation point, beyond which hardly any new votes/comments are received. Let time $t$ be $t=0$ when the story $s$ is submitted. If for this story $s$, $n_s(t)$ is the number of votes at time $t$ and $M$ is the total number of votes at saturation, then the probability distribution of votes with time, $p(t,s)$, is given by
\begin{equation}
n_s(t)=M\cdot p(t,s).  
\end{equation}
Using non-linear least curve data fitting, we fit the observed rebroadcasting behavior $p(t,s)$ with an exponential curve given below:
\begin{equation}
 \bar {p}(t,s)=h_1\cdot (1-e^{-h_2 t})
\label{eq:curvefitting}
\end{equation}

For a story $s$,  the expected number of votes at time $t$ $E[N(t)]$ , in terms of  probability of no votes at time $t$,  $\bar p_0(t,s)$ is defined by  $E[N(t)]=M*(1-\bar p_0(t,s))=M*(\bar p(t,s))$.

We see that in most cases, for each story/video/download considered, this curve gives a very good fit with the observed probability density. To illustrate this, consider an example story sampled arbitrarily for Reddit, Digg and Youtube, shown in Figures \ref{fig:Reddit} (a) \ref{fig:Digg}(a),\ref{fig:Youtube} (a) respectively. The blue curves in Figures \ref{fig:Reddit} (a) \ref{fig:Digg}(a),\ref{fig:Youtube} (a) shows the probability of votes/comments with time on an example story/video in Reddit, Digg and Youtube respectively. The curves in red are the exponential fits (Equation \ref{eq:curvefitting}). We see that these fits very well explain the observed phenomena as demonstrated by the high values of $R^2$ ($R^2>0.9$).

Similarly, in Figures \ref{fig:Reddit} (b) \ref{fig:Digg}(b),\ref{fig:Youtube} (b), the plots in blue, represent the mean probability density of voting/commenting with time averaged over all stories considered for each of the social media datasets. Again the exponential plots in red successfully explain the blue curves.

In the case of server downloads, the authors postulated that a functional curve of the form  $N(t)=\frac{a}{1-b}t^{1-b}+c*t$ (in green) best describes ($R^2=0.988$) the evolution of the number of downloads or hits on the server with time (in blue). We note that this  functional curve has a component increasing linearly with time. However, we have observed that most social phenomena and the counting process associated with them saturate with time. The downloads on a research paper wears down with time and probably does no go on forever linearly. Thus these observations only capture the transient state of the evolution of downloads. Instead, we see how well our universal curve describes the transient stage of the evolution.
 We digitized the graph in Figure 1 of that paper. We show that the universal function $h_1(1-e^{-h_2 t})+h_3$ in red well describes the evolution of the number of votes with time ($R^2=0.98$).

Based on these observations, is it then possible to make interesting deductions about a probable stochastic process that could be used to explain them?

Poisson process is a stochastic process often used to count and model the number of events occurring with time. In our case the occurring events are the votes, comments or downloads. However, if we assume the observed process in Poisson, then the probability of votes/comments would change linearly  with time as we show in the Appendix. Contrary to that, we observe an exponential change in the probability of voting with time.  One plausible reason for the Poisson process not being successful in describing the phenomena is because it assumes independence of events.  However, the probability of a person commenting on  a video might depend on a previous occurrence of commenting (maybe by his friend who initiated him to the video). Thus, the observable events like each vote/ comment/download might not be completely independent events. Rather they might be caused/correlated with each other leading to a hierarchy of correlation or causation between occurrences.
This is a signature of an ultradiffusive process described above.


 In the next section we show how the expected number of votes and observed universal curve can be explained using ultradiffusion.

\subsection{Ultrametricity of Response of Information Relaxation}
If the process is indeed ultradiffusive, then there must exist an ultrametric space on which distances between occurrences is defined. We discover such an ultrametric space.

Let a time-series $X_{t_0},X_{t_1}\cdots X_{t_n}$ ($t_i<t_j$, if $i<j$) be associated with event $e$. 
 Let $t_n$ be the bounding time on observations or the current time.
The distance between two events $X_{t_i}$ and $X_{t_j}$ be given by
\begin{equation}
d(X_{t_i},X_{t_j}) = \begin{cases}
  |max(t_n-t_i, t_n-t_j)| , & \text{if } i \ne j, \\
  0, & \text{otherwise}.
\end{cases}
\label{eq:ultrametric}
\end{equation}
   A function so defined satisfies the \emph{ultrametric distance metric properties}  defined in the previous section and the associated space is ultrametric. 
   
\begin{figure}[htbp] %
   \centering
   \begin{tabular}{c}
 \includegraphics[width=0.4\textwidth]{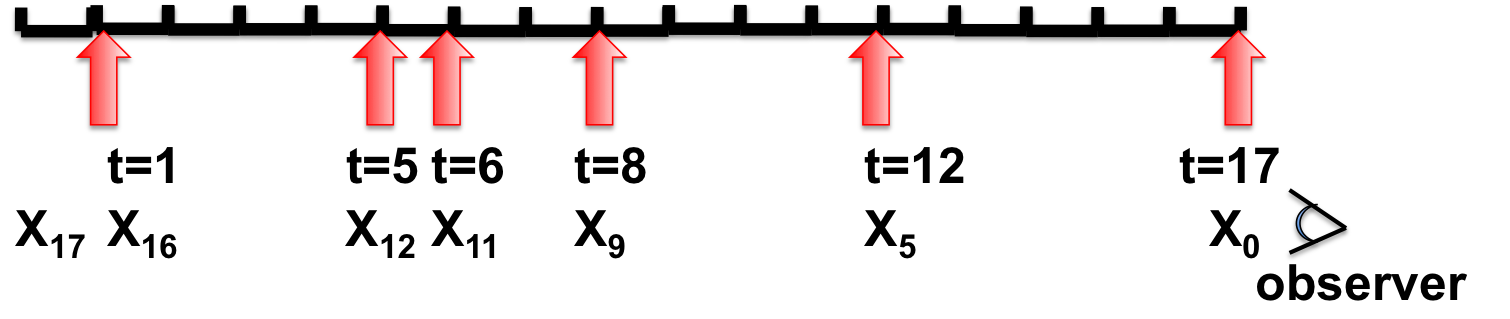}  \\
(a)
   \end{tabular}
   \caption{A timeline of events (rebroadcasts). The red arrows in the timeline show when the events occur at  times $t=1,5,6,8,12,17$. Matrix in Equation \ref{eq:timeline} shows the distance between event in the $i^{th}$ row and the $j^{th}$ column.}
   \label{fig:timeline}
\end{figure}

   Let us illustrate this distance further with a toy example. Consider a story s whose voting trace is shown in Figure \ref{fig:timeline}. In other words votes(events) occur at times $t=1,5,6,8,12$ and $17$. The time of observation is $t=17$. With each occurrence of a vote at time $t_i$, we associate a random variable $X_{t_n-t_i}$ where $t_n$ is time of observation or current time.

The distance between the different  voting occurrences is shown by the matrix in Equation \ref{eq:timeline}. For instance, as we see in Equation  \ref{eq:timeline}, the distance between the event occurring at time $t=6$, $X_{11}$ and that at time $t=12$, $X_{5}$ is given in the row corresponding to $X_{11}$ and the column corresponding to $X_5$. This distance can be seen to be $12$ from this look up matrix. The distance between $X_{17}$ (state when no votes occur) and $X_6$ is $6$.

\begin{eqnarray}
\bordermatrix{
&X_{0}    & X_{5}    & X_{9}     & X_{11} & X_{12} &X_{16} & X_{17}    \cr
         X_{0}     & 0     & 17     & 17 & 17 &17 &17 &17   \cr
    X_{5}     & 17     & 0     & 12 & 12 &12 &12 &12    \cr
    X_{9}      & 17     &12     & 0 & 8 &8 &8 &8     \cr
  X_{11}         & 17     & 12    & 8 & 0 &6 &6 &6     \cr
X_{12}       & 17     & 12     & 8 & 6 &0 &5 &5     \cr
X_{16}       & 17    & 12     & 8 & 6 &5 &0 &1    \cr
X_{17}        & 17     & 12     &8 & 6 &5 &1 &0     \cr
            }
\label{eq:timeline}
\end{eqnarray}

The existence of many time-scales in rebroadcasting behavior and our ability to construct an ultrametric space to represent the rebroadcasting behavior, leads us to speculate that response to information is an ultradiffusive process.


To simulate ultradiffusion, we first need to have a transition matrix  as in  Equation \ref{eq:diffusion}.
We exposed the ultrametric space for information response and the associated ultrametric distance above which we use to create the transition matrix  $\epsilon$ using transition rates
$\epsilon_{X_i, X_j}=e^{(-\mu d(X_i, X_j))}$.
The eigen spectrum of this transition matrix is provided in the Appendix.

\subsection{Information Relaxation is Ultradiffusive}
Using the transition matrix defined above and Equation \ref{eq:diffusion}, we can calculate the probability of being in any state at time $t$. 
Let the starting vector be $P_{X_{t_i}}(0)=\chi_{X_{t_i}}$ where $\chi_{X_{t_i}}$ is a characteristic function  (i.e. $P_{X_{t_i}}(0)$ is a vector whose  element corresponding to $X_{t_i}$ is $1$ and it is zero elsewhere). The autocorrelation function is $P_{X_{t_i}}(t)$ which corresponds to state $X_{t_i}$. 

$P_{X_{t_N}}(t)$ defines the probability of being in state $X_{t_N}$ at time $t$. If $X_{t_N}$ is analogous to the non-rebroadcasting (voting, commenting) state (like $X_{17}$ in Figure \ref{fig:timeline} ) and $P_{X_{t_N}}(t)$ is analogous to the probability of not rebroadcasting till time $t$, then taking $M$ as the maximum number of rebroadcasts possible,  $M*(1- P_{X_{t_N}}(t))$ is the expected number of rebroadcasts at time $t$. Since there are no votes initially, therefore the probability of not voting $i.e.$, $P_{X_{t_N}}(0)=1$ and $P(0)=\chi_N$.

If the number of states in the transition matrix are finite,  we know that the decay of the autocorrelation function  is  exponential \cite{BachasHuberman}.

Specifically for transition matrix defined using ultrametric distance in Equation \ref{eq:ultrametric} (and whose eigen spectrum is discussed in details in the Appendix),  we have
$P_{X_{t_i}}(t)=\sum_{j=1}^{t_N} e^{(\lambda(j) t)}V^2_i(j)$

$P_{X_{t_i}}(t)=\frac{i-1}{i}*e^{ -t( (i-1)e^{(-\mu(i-1))}+\sum_{k=i}^{t_N} e^{(-\mu (k-1)))}) }+$

$\sum_{j=i+1}^{t_N} \frac{1}{j.(j-1)} e^{ -t( (j-1)e^{(-\mu(j-1))}+\sum_{k=j}^{t_N} e^{(-\mu (k-1)))} } +\frac{1}{N}$

Therefore, $P_{X_{t_N}}(t)=\frac{t_N-1}{t_N}*e^{ -t( t_N\dot e^{(-\mu(t_N-1))}}+\frac{1}{t_N}$

and the expected number of rebroadcasts after time t is   $E(R(t))=M*(1- P_{X_{t_N}}(t))=\frac{1}{t_N}*(1-e^{ -t( t_Ne^{(-\mu (t_N-1))}})$.

 

\section{Experiments}
We hypothesize the ultrametricity of information response and that relaxation of information response is ultradiffusive. When the information is a story shared on Reddit or Digg or a video shared on Youtube, we claim its growth in popularity can be rationalized using a ultradiffusion based stochastic process.


To verify this claim, for each story/news/video in the social media we simulate its spread using an ultradiffusion process.
The  transition matrix describing the ultradiffusion process is determined by the ultrametric distance $d({X_{t_i},X_{t_j}})$ between events $X_{t_i}$ and $X_{t_j}$ are defined as in Equation \ref{eq:ultrametric}. The transition rates between $X_{t_i}$ and $X_{t_j}$ are given by $e^{-\mu {d({X_{t_i},X_{t_j}})} }$.  However, prior to simulation, we need to determine the values of $t_N$ (the bounding time on observations) and $\mu$ (the scaling factor) for the ultradiffusion process to be simulated. We learn these variables from the observed data on rebroadcasting behavior.


We use $h_1$ and $h_2$ in Equation \ref{eq:curvefitting} to  learn the values of $t_N$ and $\mu$.
The length of the hierarchical tree $t_N$ is given by $t_N=\frac{1}{1-h_2}$. The decay parameter $\mu$ is deduced as $\mu=log(\frac{t_N}{h_1})/(t_N-1)$.

Once we have deduced $t_N$ and $\mu$, we can then use these values to calculate the transition matrix $\epsilon$ encoding the ultradiffusion process. We assume that the ultradiffusion process starts at state $N$ $i.e$ $P_{X_{t_N}}(0)=\chi_{X_{t_N}}$ in Equation \ref{eq:diffusion}. As explained above, $P_{X_{t_N}}(t)$, gives the probability of not voting which is $1$ at $t=0$.  We then simulate the ultradiffusion process using the learned parameters $t_N$ and $\mu$. The rebroadcasting behavior obtained by simulating ultradiffusion is then given by
\begin{eqnarray}
\hat{p}(t,s)&=&1-P_{X_{t_N}}(t) \nonumber \\
&=&\frac{1}{t_N}(1-e^{ -(t {t_N}e^{(-\mu(t_N-1))}})
\end{eqnarray} 

We plot the observed probability curve $p(t,s)$ in blue, the fitted probability curve $\bar p(t,s)$ (in red) and the probability curve obtained from simulated ultradiffusion $\hat p(t,s)$ (in black) with time $t$ for every story/news/video $s$ investigated in the different social media websites.

To get the overall trend of information spread in the different social media, we average the probability of voting across different stories/news/videos for each social media website.  Then we simulate an ultradiffusion process generating this overall trend. In other words,  just like for individual stories, for each social media, we fit the mean probability of voting with an exponential curve. We use this fit to determine the $t_N$ and $\mu$ of an ultradiffusion process. We then calculate the mean probability of voting with time using this ultradiffusion process  and compare it with the observed mean probability of voting with time.

We evaluate the goodness of fit using $R^2$ which measures how successful the fit is in explaining the variation of the data. High value of $R^2$ indicates a good fit.

 Irrespective of the social media studied, we observe that for most stories/news/videos $s$, the generated rebroadcasting behavior by simulating ultradiffusion represented by $\hat p(t,s)$ closely agrees with the observed rebroadcasting behavior given by $p(t,s)$. We also observe that the ultradiffusion process seems to explain the overall rebroadcasting behavior in the different social media satisfactorily. We demonstrate these results in greater details in the next section.
\section{Results and Evaluation}
In this section we present our observations of how the cumulative probability density of votes (in Reddit and Digg) and comments (in Youtube) changes with time. For most of the stories the rebroadcasting behavior calculated using the ultradiffusion model fits very well with the observed probability of voting with time. We calculate the probability of voting through ultradiffusion simulation for each individual story in Reddit, Digg and Youtube as shown in Figures \ref{fig:Reddit}(a),\ref{fig:Digg}(a) and \ref{fig:Youtube}(a). We also, calculate the mean probability of voting over a collection of stories in each of these social media as as shown in Figures \ref{fig:Reddit}(b),\ref{fig:Digg}(b) and \ref{fig:Youtube}(b).

 We see a high correspondence between the actual and the simulated rebroadcasting behavior irrespective of the social media considered.

\subsection{Reddit}
We investigate the rebroadcasting behavior for individual stories in the Reddit dataset.
For a sample story, the actual probability of voting with time (in blue), its curve fitting (in red) and the probability of voting calculated from an ultradiffusion simulation (in black) is shown in Figure\ref{fig:Reddit} (a).   Multiplying this probability with the maximum number of votes gives the expected number of votes at time any time $t$ for that story. As shown in Figure \ref{fig:Reddit} (b), we also calculate the mean probability of voting on Reddit (in blue), the best exponential fit (in red) and the predicted probability of voting using ultradiffusion (in black).
We evaluate how successful ultradiffusion is in explaining the probability of voting (both for each individual stories and for the average over all stories on Reddit) using $R^2$. In most cases we get a very high value of $R^2$ as demonstrated in Figure\ref{fig:Reddit}. This means that ultradiffusion is very successful in describing rebroadcasting behavior in Reddit.

\subsection{Digg}

In Figure\ref{fig:Digg} (a), we give an example of how the rebroadcasting behavior (digging/voting behavior)  on a story changes with time. The probability of voting with time is shown in blue, the curve fitted is shown in red and the rebroadcasting behavior derived from the simulated ultradiffusion process is shown in black. Again, for the given story $s$, if the maximum votes/diggs received is $M$, then $M$ times the probability of voting at time $t$, $p(t,s)$, (an example of which is shown in Figure \ref{fig:Digg} ) gives the expected number of votes that story $s$ received that time $t$. Just like in Figure \ref{fig:Reddit} (b), Figure \ref{fig:Digg} (b) gives the observed mean probability of voting over all stories in Digg (in blue), the exponential fitted curve (in red) and the predicted probability of voting using ultradiffusion in black. Again high value of $R^2$ both for individual stories and for the average of of all stories on Digg ($R^2=0.95$) as shown in  Figure \ref{fig:Digg} demonstrates that ultradiffusion explains with high accuracy the rebroadcasting behavior on Digg.

\subsection{Youtube}
The actual probability commenting with time (in blue), its curve fitting (in red) and the rebroadcasting behavior simulated using an ultradiffusion process (in black)  on an arbitrarily chosen video on Youtube is given in Figure\ref{fig:Youtube}(a). Similar to the rebroadcasting activities in Digg and in Reddit, note that when this probability is multiplied by the the total number of comments at saturation, we get the expected  number of comments at time $t$. Just like in Reddit and Digg, we also measure the mean probability of commenting over all Youtube videos collected (shown in blue in Figure \ref{fig:Youtube} (b)), the corresponding exponential fit (in red) and the predicted mean probability of voting using ultradiffusion (which is shown by the black curve in the Figure \ref{fig:Youtube} (b)). Again on evaluating the goodness of fit, we get a high value of $R^2$, suggesting that ultradiffusion very well explains the data.

\subsection{Server Downloads}

We observe in Figure \ref{fig:server}, that ultradiffusion can also be successfully describe the evolution of the number of downloads from a server with time. The observed probability of downloads with time is shown in blue and the probability predicted using ultradiffusion is shown in black . Again we see a very high correspondence between the actual observations of the counting process and that predicted by ultradiffusion($R^2=0.97$)

We see that in each of the four datasets  the predicted  probability of votes  by simulating ultradiffusion process (both for individual stories and the overall social)  is very close to the observed probability of voting. Therefore, we claim that the
rebroadcasting behavior generated by the ultradiffusion process is very similar to the actual rebroadcasting behavior observed.
\remove
{
\section{Discussion}

\emph{Poisson process} has often been used to describe human behavior. However, an important assumption in Poisson processes is the independence of occurring events. As we show in the Appendix, the expected number of votes at time $t$ if the process would have been Poisson is $M \rho t$. Here $\rho$ is the rate of arrival. We note that in the ultradiffusion process, when  the scaling factor $\mu$ is very large, then the ultrametric distance between different states is very large and the probability of transition between different states is small. In such a case, the states tend to be independent of each other and the ultradiffusion process reduces to the Poisson process. Thus the Poisson process is a limiting case of the ultradiffusion process.
}
\section{Conclusion}
We study the relaxation of human response to information on online social networks.   For  of news, stories or videos, this response is in terms of votes/comments that the story/video receives over time. We discover a universal function describing the observed counting process regardless of the social media considered; be it the probability of voting on a story on Reddit or Digg or the probability of commenting on a video on Youtube with time.  The parameters of this function can be learned from the dataset under consideration.

We notice that the inter-arrival time between votes/comments is not homogeneous and there exists a hierarchy of time scales. Some consecutive votes occur almost simultaneously whereas others might occur with along interval within them. This leads us to postulate that these occurrences of these events might not necessarily be completely independent  with some events being correlated with each other as  is  signature of an ultradiffusive process.  The universal function observed can indeed be well described using ultradiffusion. On the other hand, it cannot be derived using the Poisson process.


We know that for  transition matrices corresponding to finite trees, the decay of the autocorrelation is \emph{exponential} with time.
However, if the hierarchical tree is infinite, there is a \emph{power law} decay of the autocorrelation function with time (\emph{cf.} Appendix).

Our ability to model this universal counting process gives us the power to predict the future evolution of the rebroadcasts over time.
Future work includes predicting the observable behavior in other social networks and disease spread.

\bibliographystyle{abbrv}
\bibliography{ultrametricity}

\begin{thebibliography}{10}

\bibitem{BachasHuberman}
C.~P. Bachas and B.~A. Huberman.
\newblock Complexity and ultradiffusion.
\newblock {\em Journal of Physics A: Mathematical and General}, 20(14):4995,
  1987.

\bibitem{Bachas87}
C.~P. Bachas and W.~F. Wolff.
\newblock Percolation and the complexity of games.
\newblock {\em Journal of Physics A: Mathematical and General}, 20(1):L39,
  1987.

\bibitem{Backstrom06}
L.~Backstrom, D.~Huttenlocher, J.~Kleinberg, and X.~Lan.
\newblock Group formation in large social networks: membership, growth, and
  evolution.
\newblock In {\em Proceedings of the 12th ACM SIGKDD international conference
  on Knowledge discovery and data mining}, KDD '06, pages 44--54, New York, NY,
  USA, 2006. ACM.

\bibitem{Bailey}
N.~T.~J. Bailey.
\newblock {\em Mathematical Theory of Infectious Diseases}.
\newblock Oxford University Press, 1975.

\bibitem{BakshysocN}
E.~Bakshy, I.~Rosenn, C.~Marlow, and L.~Adamic.
\newblock The role of social networks in information diffusion.
\newblock In {\em Proceedings of the 21st International Conference on World
  Wide Web}, WWW '12, pages 519--528, New York, NY, USA, 2012. ACM.

\bibitem{Barabasi05nature}
A.-L. Barab\'{a}si.
\newblock {The origin of bursts and heavy tails in human dynamics}.
\newblock {\em Nature}, 435:207--211, May 2005.

\bibitem{Castellano}
C.~Castellano and R.~Pastor-Satorras.
\newblock Thresholds for epidemic spreading in networks.
\newblock {\em Phys. Rev. Lett.}, 105:218701, Nov 2010.

\bibitem{CentolaD}
D.~Centola.
\newblock {The Spread of Behavior in an Online Social Network Experiment}.
\newblock {\em Science}, 329(5996):1194--1197, Sept. 2010.

\bibitem{Cha2010}
M.~Cha, H.~Haddadi, F.~Benevenuto, and K.~P. Gummadi.
\newblock Measuring user influence in twitter: The million follower fallacy.
\newblock In {\em in ICWSM �10: Proceedings of international AAAI Conference
  on Weblogs and Social}, 2010.

\bibitem{Cui13}
P.~Cui, S.~Jin, L.~Yu, F.~Wang, W.~Zhu, and S.~Yang.
\newblock Cascading outbreak prediction in networks: a data-driven approach.
\newblock In {\em Proceedings of the 19th ACM SIGKDD international conference
  on Knowledge discovery and data mining}, KDD '13, pages 901--909, New York,
  NY, USA, 2013. ACM.

\bibitem{GomezRodriguez13}
M.~Gomez~Rodriguez, J.~Leskovec, and B.~Sch\"{o}lkopf.
\newblock Structure and dynamics of information pathways in online media.
\newblock In {\em Proceedings of the sixth ACM international conference on Web
  search and data mining}, WSDM '13, pages 23--32, New York, NY, USA, 2013.
  ACM.

\bibitem{Versteeg11icwsm}
{Greg {Ver Steeg}}, R.~Ghosh, and K.~Lerman.
\newblock What stops social epidemics?
\newblock In {\em Proceedings of 5th International Conference on Weblogs and
  Social Media}, 2011.

\bibitem{Gruhl04}
D.~Gruhl, R.~Guha, D.~Liben-Nowell, and A.~Tomkins.
\newblock Information diffusion through blogspace.
\newblock In {\em Proceedings of the 13th International Conference on World
  Wide Web}, WWW '04, pages 491--501, New York, NY, USA, 2004. ACM.

\bibitem{Hethcote}
H.~W. Hethcote.
\newblock An immunization model for a heterogeneous population.
\newblock {\em Theor. Pop. Biol.}, 14:338--349, 1978.

\bibitem{Kerzberg}
B.~A. Huberman and M.~Kerzberg.
\newblock Ultradiffusion- the relaxation of hierarchical systems.
\newblock {\em Journal of Physics A: Mathematical and General}, 18(6):4470,
  1985.

\bibitem{Sornette}
A.~Johansen and D.~Sornette.
\newblock Download relaxation dynamics on the \{WWW\} following newspaper
  publication of \{URL\}.
\newblock {\em Physica A: Statistical Mechanics and its Applications},
  276(1�2):338 -- 345, 2000.

\bibitem{Kephart}
J.~Kephart and S.~White.
\newblock Directed-graph epidemiological models of computer viruses.
\newblock In {\em Research in Security and Privacy, 1991. Proceedings., 1991
  IEEE Computer Society Symposium on}, pages 343--359, 1991.

\bibitem{Kwak10}
H.~Kwak, C.~Lee, H.~Park, and S.~Moon.
\newblock What is twitter, a social network or a news media?
\newblock In {\em Proceedings of the 19th international conference on World
  wide web}, WWW '10, pages 591--600, New York, NY, USA, 2010. ACM.

\bibitem{lermanGhosh}
K.~Lerman and R.~Ghosh.
\newblock Information contagion: an empirical study of spread of news on digg
  and twitter social networks.
\newblock In {\em Proceedings of 4th International Conference on Weblogs and
  Social Media (ICWSM)}, May 2010.

\bibitem{viraldynamics}
J.~Leskovec, L.~A. Adamic, and B.~A. Huberman.
\newblock The dynamics of viral marketing.
\newblock {\em ACM Trans. Web}, 1(1), May 2007.

\bibitem{Malmgren08}
R.~D. Malmgren, D.~B. Stouffer, A.~E. Motter, and L.~A.~N. Amaral.
\newblock A poissonian explanation for heavy tails in e-mail communication.
\newblock {\em Proceedings of the National Academy of Sciences},
  105(47):18153--18158, 2008.

\bibitem{NewmanEpidemic}
M.~E.~J. Newman.
\newblock {\em Physical Review E}, (1):016128, July.

\bibitem{Vespignani2001}
R.~Pastor-Satorras and A.~Vespignani.
\newblock Epidemic spreading in scale-free networks.
\newblock {\em Phys. Rev. Lett.}, 86:3200--3203, Apr 2001.

\bibitem{Prakash}
B.~A. Prakash, D.~Chakrabarti, M.~Faloutsos, N.~Valler, and C.~Faloutsos.
\newblock Got the flu (or mumps)? check the eigenvalue!
\newblock 2010.

\bibitem{Romero2011}
D.~M. Romero, W.~Galuba, S.~Asur, and B.~A. Huberman.
\newblock Influence and passivity in social media.
\newblock In {\em Proceedings of the 20th international conference companion on
  World wide web}, WWW '11, pages 113--114, New York, NY, USA, 2011. ACM.

\bibitem{Gabor}
G.~Szabo and B.~A. Huberman.
\newblock Predicting the popularity of online content.
\newblock {\em Commun. ACM}, 53(8):80--88, Aug. 2010.

\bibitem{Watts}
D.~J. Watts, P.~S. Dodds, and M.~E.~J. Newman.
\newblock Identity and search in social networks.
\newblock {\em Science}, 296(5571):1302--1305, 2002.

\bibitem{weng:virality}
L.~Weng, F.~Menczer, and Y.-Y. Ahn.
\newblock {Virality Prediction and Community Structure in Social Networks}.
\newblock {\em Scientific Reports}, 3, Aug. 2013.

\bibitem{wu04information}
F.~Wu, B.~A. Huberman, L.~A. Adamic, and J.~R. Tyler.
\newblock Information flow in social groups.
\newblock {\em Physica A: Statistical and Theoretical Physics},
  337(1-2):327--335, June 2004.

\end{thebibliography}


\begin{thebibliography}{1}
\bibitem{BachasHuberman}
C.~P. Bachas and B.~A. Huberman.
\newblock {Complexity and ultradiffusion}.
\newblock {\em Journal of Physics A: Mathematical and General}, 20(14):4995,
  Oct. 1987.
\bibitem{Kerzberg}
 B.~A. Huberman and M. Kerzberg
\newblock {Ultradiffusion:the Relaxation of Hierarchical Systems}.
\newblock {\em Journal of Physics A: Mathematical and General}, 18:L331,
  Jan. 1985.
\bibitem{viraldynamics}
Jure Leskovec, Lada~A. Adamic, and Bernardo~A. Huberman.
\newblock The dynamics of viral marketing.
\newblock {\em ACM Trans. Web}, 1(1), May 2007.
\bibitem{cha2010measuring}
M.~Cha, H.~Haddadi, F.~Benevenuto, and K.P. Gummadi.
\newblock Measuring user influence in twitter: The million follower fallacy.
\newblock In {\em 4th International AAAI Conference on Weblogs and Social Media
  (ICWSM)}, 2010.
\bibitem{lermanGhosh}
Lerman, K., and Ghosh, R.
\newblock Information contagion: An empirical study of the spread of news on
  digg and twitter social networks.
\newblock In {\em 4th International AAAI Conference on Weblogs and Social Media
  (ICWSM)}, 2010.

\end{thebibliography}
\section{Appendix}
\subsection{ Poisson Process}
Let $\rho$ be the rate of arrival of a Poisson process.  The probability of the number of votes at time $t$ being $k$ is given by 
\begin{equation}
P(N(t)=k)= (e^{-\rho t}{\rho t}^k)/k!
\end{equation}

(Since $ \sum_{k=0}^{\infty} (e^{-\rho t}{\rho t}^k)/k!=1)$

We get:
\begin{equation}
E[N(t)]=\sum_{k=0}^{\infty}k (e^{-\rho t}{\rho t}^k)/k!=\rho t 
\end{equation}

If $T_0$ is the time at saturation, then the probability of an event occurring is given by $\frac{t}{T_0}$. Hence the probability of an event occurring varies linearly with time.
\subsection{Ultradiffusion}
\cite{BachasHuberman}  gives the exact solution for the autocorrelation at state $L$ which is taken as a leaf of a hierarchical tree:
\begin{equation}
P_{L}(t)=\frac{1}{N}+\sum_{n=1}^{root}(\frac{1}{N_{L_{n-1}}}-\frac{1}{N_{L_{n}}}) e^{(-\frac{t}{\tau_{L_n}})}
\label{eq:5}
\end{equation}
Here $N$ is the number of leaves of the tree, $L_n$ is the node in the $n^{th}$ level of the shortest path from leaf $L$ to the root. The level of the leaf $L$ is minimum ($n=1$) and that of the root is maximum. $N_{L_n}$ are the number of final descendants or leaves of node $L_n$.  $\frac{1}{\tau_{L_n}}$ is given by:
\begin{equation}
\frac{1}{\tau_{L_n}}=N_{L_n}e^{(-h_{L_n})}+\sum_{i=2}^{root}(N_{L_i}-N_{L_{i-1}})e^{(-h_{L_i})}
\label{eq:3}
\end{equation}
Here $h_{L_i}$ is the height of node $L_i$. The height of the leaves is the minimum and that of the root is the maximum.
Therefore $\alpha=\frac{1}{N}$, $\beta_n=\frac{1}{N_{L_{n-1}}}-\frac{1}{N_{L_{n}}}$ and $\gamma_n=\frac{1}{\tau_{L_n}}$ in Equation \ref{eq:4}.

\subsection{Power Laws}
If we consider a hierarchical tree, in which every node produces $b$ offsprings and $\Delta h$ is the height between two levels,  then the silhouette of the tree is given by 
\begin{equation}
s=\frac{1}{\Delta h} log b
\end{equation}
If $s<1$ the autocorrlation function averaged over all initial conditions is given by 
\begin{eqnarray}
\bar P(t)&=&\sum_{m=1}^{\infty} (b-1)b^{-m} e^{(-t{(b e^{-\Delta h})}^m \frac{e^{\Delta h}-1}{e^{\Delta h}-b})}\nonumber \\
&=&D t^{-v}
\end{eqnarray}
Here  $v=\frac{s}{1-s}$ and

 $D=\Gamma v {(\frac{e^{\Delta h}-1}{e^{\Delta h}-b})}^{-v} \frac{b-1}{log b} v$
 
\cite{BachasHuberman} also show that uniformly random trees have the same dynamic exponent(silhouette)   as completely ordered uniform trees.

\subsection{Eigen Spectrum of the Transition Matrix}
We construct a transition matrix for a time series of events $X_{t_1},X_{t_2}\cdots X_{t_N}$ ($t_i<t_j$, if $i<j$)  whose ultrametric distance metric  is defined by $d(X_{t_i},X_{t_j})=|max(t_N-t_i, t_N-t_j)|$ if $i \ne j$ and 0 otherwise. For this matrix we construct a hierarchical tree in lines of \cite{BachasHuberman}  and  calculate the normalized eigenvalues using the characteristic function
\[X_{t_i}(j)=\left\{\begin{array}{c c}
1 & \quad \text{if $t_i$ is a descendant of $t_j$}\\
0 & \quad \text{otherwise}
\end{array}\right.\]
$N_y$ are the number of descendants or leaves generated by $y$. We take $ y$ as a son of $y_1$.
We note that $v_{t_i}(y)$ is a basis for the subspace the eigenvectors span, one for each son of $y$.
$v_{t_i}(y)=\frac{1}{N_y}X_{t_i}(y)-\frac{1}{N_{y_1}}X_{t_i}(y_1)$. 

Taking the degeneracy and linear independence under consideration, we derive analytically that if there are $t_N$ leaves, then the eigenvectors are given by:
\[\bar V_{t_i}(j) = \left\{ 
  \begin{array}{l l}
    1/t_N & \quad \text{if $j=1$ $\forall$ $i$ }\\
    1/(j-1) -1/(j) & \quad \text{if $1<j \le t_N$ and $i<j$} \\
    -1/(j) & \quad \text{if $1<j \le t_N$ and $j=i$} \\
    0& \quad \text{if $1<j \le t_N$ and $i>j$} \\
  \end{array} \right.\]
The normalized eigenvectors are given by
\[ V_i(j) = \left\{ 
  \begin{array}{l l}
    1/\sqrt{t_N} & \quad \text{if $j=1$ $\forall$ $i$ }\\
    1/\sqrt{(j-1).(j)} & \quad \text{if $1<j \le t_N$ and $i<j$} \\
    -\sqrt{j-1/j} & \quad \text{if $1<j \le t_N$ and $j=i$} \\
    0& \quad \text{if $1<j \le t_N$ and $i>j$} \\
  \end{array} \right.\]

The corresponding eigenvalues are given by:
\[ \lambda(j) = \left\{ 
  \begin{array}{l l}
    0& \quad \text{if $j=1$ }\\
  -( (j-1)e^{(-\mu(j-1))}+\\
  \sum_{i=j}^{t_N} e^{(-\mu (i-1)))}& \quad \text{if $1<j \le t_N$} \\
  \end{array} \right.\]
\remove
{

}
\end{document}